\documentclass[aip,jcp,numerical,preprint]{revtex4-1}
\usepackage[english]{babel}
\usepackage{graphicx}
\usepackage{color}
\begin{document}
\title{In a search for a shape maximizing packing fraction for two-dimensional random sequential adsorption}

\author{Micha\l{} Cie\'sla}
 \email{michal.ciesla@uj.edu.pl}
\author{Grzegorz Paj\c{a}k}
\affiliation{M.\ Smoluchowski Institute of Physics, Department of Statistical Physics, Jagiellonian University, \L{}ojasiewicza 11, 30-348 Krak\'ow, Poland.}

\author{Robert M. Ziff}
 \email{rziff@umich.edu}
\affiliation{Center for the Study of Complex Systems and Department of Chemical Engineering, University of Michigan, Ann Arbor MI 48109-2136 USA.}

\date{\today}
\begin{abstract}
Random sequential adsorption (RSA) of various two dimensional objects is studied in order to find a shape which maximizes the saturated packing fraction. This investigation was begun in our previous paper [Cie\'sla \textit{et al., Phys. Chem. Chem. Phys.}, \textbf{17}, 24376 (2015)], where the densest packing was studied for smoothed dimers. Here this shape is compared with a smoothed $n$-mers, spherocylinders and ellipses. It is found that the highest packing fraction out of the studied shapes is $0.58405 \pm 0.0001$ and is obtained for ellipses having long-to-short axis ratio of $1.85 \pm 0.07$.
\end{abstract}
\maketitle
\section{Introduction}
Random sequential adsorption (RSA) is a conceptually easy procedure to randomly deposit objects on a hypersurface which, for simplest cases, is flat and homogeneous. Historically, a one-dimensional version of the RSA problem occurred first in 1939 \cite{Flory1939}, where interactions between attachments to a discrete polymer line was studied. Then, in 1959, A. R\'enyi found analytically the saturated random packing density for one-dimensional continuum RSA by solving the so-called car parking problem \cite{Renyi1958}. Later two-dimensional (2D) RSA became a very successful approach for modeling monolayers in the process of irreversible adsorption \cite{Feder1980,Evans1993,Adamczyk2012}. A sketch of the 2D RSA algorithm for anisotropic shapes is as follows: (i) the position and orientation of the virtual particle is drawn according to the probability distribution that reflects properties of an underlying substrate --- when a surface is homogeneous, this probability distribution will be uniform; (ii) next it is tested if the virtual particle overlaps or intersects with any of particles already added to the packing; (iii) if not, the particle is added to the packing, otherwise, if there is an overlap, it is abandoned. Such  attempts should proceed until the packing is saturated, that is, to the final case when there is no space for any virtual particle on a substrate. In practice, because of substantial slowing down, the simulation is terminated when the probability of successfully adding a virtual particle is sufficiently small, with further extrapolation to the saturated packing. Details on the extrapolation method and on when to stop the algorithm at certain accuracy are given in the following sections of this article. 

In this study we are searching  for the shape that gives densest saturated random packing. The relation between particle shape and packing density has been intensively studied in the context of  the related problem of random close packings (RCP), where particles are tightly packed in a jammed configuration and are touching  many of their closest neighbors \cite{Baule2014}. Interestingly, in the case of RCP, convex anisotropic shapes typically give denser packings than disks or spheres \cite{Donev2004, Man2005}, and the densest packing were found for particles (ellipsoids, spherocylinders or dimers) of long-to-short axis ratio around $1.5$ \cite{Donev2004, Faure2009, Zhao2012, Baule2013}. Recently, it has been analytically proven that sufficiently sphere-like, but anisotropic shapes pack more densely than spheres, which is in accordance with the Ulam's conjecture, which posits that spheres have the lowest optimal packing density among all convex shapes \cite{Kallus2016}. Although, in the case of RSA this is not true for elongated rectangles \cite{Vigil1990} as well as for squares \cite{Viot1990} that gives lower saturated packing fraction than disks, it is still possible that the disk is a local minimum of the packing density in some space of shapes. However, such behavior is not supported by the results obtained for regular polygons \cite{Ciesla2014star}, where packing fractions seems to be slightly below the one for disks.

In our recent report \cite{Ciesla2015} we have studied several shapes and shown that the highest packing of $0.5833$ was obtained for a smoothed dimer --- a concave shape derived from a dimer of two overlapping disks. This value is comparable to the maximum packing fraction reported earlier for ellipsoids and spherocylinders \cite{Sherwood1999,Viot1992}, but due to the numerical accuracy of these results it was impossible to determine which of these three shapes gives the highest maximal random coverage. The main aim of this study is to settle this issue. Moreover, we also find the packing fraction for shapes derived from linear polymers that are in between smoothed dimers and spherocylinders. Additionally, we have also included concepts introduced in Ref.\ \cite{Zhang2013}, which resulted in speeding up the RSA algorithm and even in obtaining saturated packings in a finite time. 

In this paper we first introduce the RSA model in detail, then we gather results for the kinetics and saturated random packings.   We also examine spatial and angular correlations in the jammed state for ellipses, followed by a discussion of the measurement errors. The article is closed by a brief summary. We provide an Appendix with explicit formulas for areas of investigated geometries. 

\section{Model}
\begin{figure}[htb]
\centerline{%
\includegraphics[width=0.9\columnwidth]{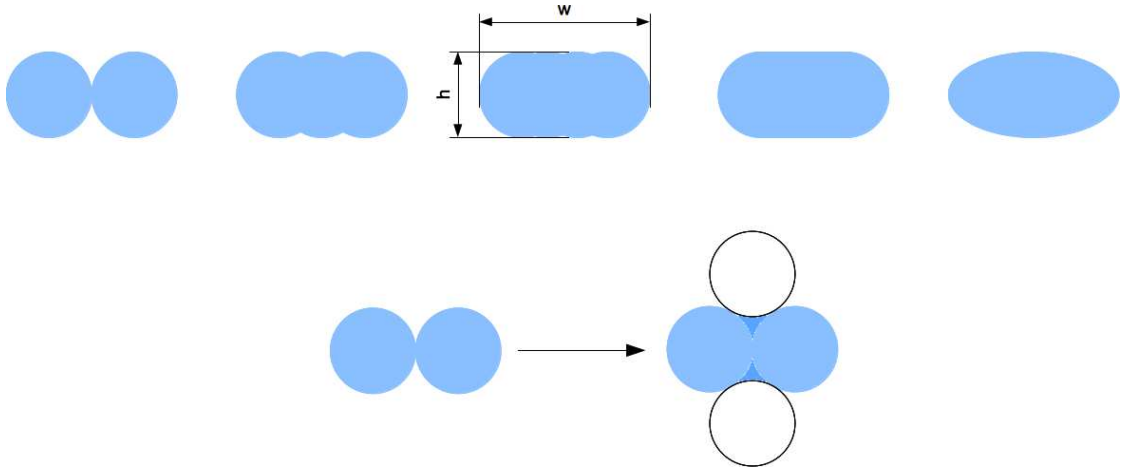}}
\caption{(Color online) The types of shapes for which saturated random packings are studied. From the left: linear polymers built of two, three and four overlapping disks of the same radius, spherocylinder and ellipse. All presented shapes have the same long-to-short axis ratio $w/h = 2$. Below it is shown how all linear polymers are smoothed.}
\label{fig:shapes}
\end{figure}

Examples of the shapes we consider are shown in Fig.~\ref{fig:shapes}. 
Linear polymers are built of identical disks. In this study we restricted ourselves to dimers, trimers, tetramers, pentamers and decamers. All these shapes were smoothed as shown in Fig.\ \ref{fig:shapes} for the case of a dimer. 
Here, we consider configuration build of smoothed shape that correspond 1-to-1 to configurations build of the unsmoothed particles, but have higher density. However, in general disks forming smoothed particles do not have to overlap or even touch themselves. But, in order to preserve the mentioned correspondence between set of disks and smoothed particles, which is useful for packing generation procedure, the distance between two closest disks centers should not be larger than $h\sqrt{2}$,  where $h$ is the disk diameter. Thus the long-to-short axis ratio for smoothed particles should not exceed $(k-1)\sqrt{2} + 1$, where $k$ is a number of disks in a particle. Besides linear polymers, we studied  spherocylinders and ellipses. The anisotropy of the shape is defined as a long-to-short axis ratio: $x = w/h$. To find the anisotropy that gives highest packing fraction we varied the parameter $x$ between $1.1$ and $2.5$. This particular interval was chosen according to results of previous studies, \emph{e.g.} \cite{Ciesla2015,Vigil1989}. Formulas for the areas covered by these shapes are collected in the Appendix. To make the comparison of packing fraction between different shapes as clear as possible, all shapes' sizes were re-scaled so that their areas are equal to $1$. For example, an ellipse of anisotropy $x$ has the short semi-axis of length $\sqrt{1/ (x\pi)}$ and the long semi-axis of length $\sqrt{x/\pi}$.  
\par
These shapes were thrown onto a square surface of a side size $1000$ with an area $S=10^6$ and periodic boundary conditions. Checking if shapes are overlapping is straightforward for the overlapping disk and sphero-cylinder cases as there it is based on simple disk-disk, disk-interval, and interval-interval intersections. In the case of ellipses the exact Vieillard-Baron criterion was used \cite{Vieillard-Baron1972}.
As the particle surface area is unity, the packing density is equal to the packing fraction:
\begin{equation}
\label{eq:nq}
\theta(t) = \frac{N(t)}{S},
\end{equation}
where $N(t)$ is a number of particles in a packing after a number of RSA iterations corresponding to time $t$ measured in the dimensionless time units 
\begin{equation}
t = \frac{n}{S},
\end{equation}
where $n$ is number of RSA algorithm steps.
\par
\begin{figure}[htb]
\centerline{%
\includegraphics[width=0.35\columnwidth]{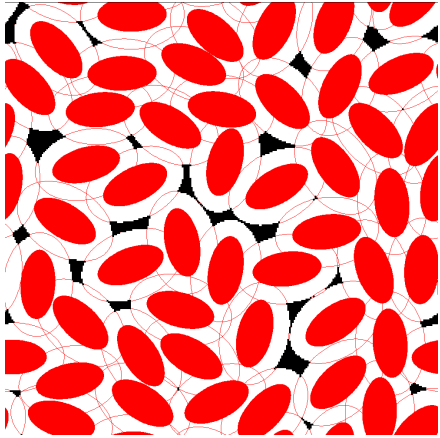}
}
\caption{(Color online) Example of ellipses random packing. An elliptic region surrounding each ellipse shows the area where it is not possible to place center of next ellipse. The only places where center of next ellipse can potentially be placed are the black regions. Here, they fill approximately 7\% of whole packing; thus, selecting random point only from these regions speeds up simulation 14 times.}
\label{fig:regions}
\end{figure}

The simulation was stopped when $t=10^6$. To improve statistics, up to $100$ independent simulations were performed for each shape. This specific values of packing size and the simulation time were chosen to ensure a desired level of numerical error of the average saturated random packing \cite{Ciesla2016}, which for our purposes should be below $0.001$. To generate random packing the modified version of RSA algorithm introduced in  Ref.\ \cite{Zhang2013} was used. The modification is that the algorithm traces the unoccupied places where subsequent particles can possibly be placed. Therefore, the random position of consecutive shape can be limited to these places. For example a center of a disk of radius $r$ cannot be closer to a boundary of any previously placed disk than $r$. This modification allows one to obtain strictly saturated packing, as such unoccupied spaces must vanish or be filled in by a disk. The simulation stops when there are no such regions. For anisotropic particles, the shape of such unoccupied regions depends on the orientation of the particle that tries to fit there. To work around this problem, we decided to exclude only areas where it is not possible to place center of particle in any possible orientation. For example, the center of subsequent spherocylinder of height $h$ cannot be closer from boundary of another spherocylinder than $h/2$ as well as the center of an ellipse cannot be closer from the other ones than its shorter semi-axis length (see Fig.\ \ref{fig:regions}).
The cost of such a solution is that there are regions in which  an anisotropic particle will never fit. For sure, there could be placed the center of disk of radius $h/2$, but we cannot determine if an anisotropic particle will fit there. Therefore, the generated packing are not saturated; however, drawing a place for the next particle only from these black regions can significantly speed up a simulation. This makes it possible to study substantially larger packings and to increase the effective number of RSA iterations compared to our previous study \cite{Ciesla2015}. Note, that one simulation step when particle position is selected only in regions of the total size $S_{reg}$ corresponds to $S/S_{reg}$ iterations of the original RSA procedure. Thus, one iteration corresponds to increase of dimensionless time by $1/S_{reg}$. During the simulations the number of particles in a packing as a function of time $N(t)$ was recorded.
\section{Results}
\begin{figure}[htb]
\centerline{%
\includegraphics[width=0.35\columnwidth]{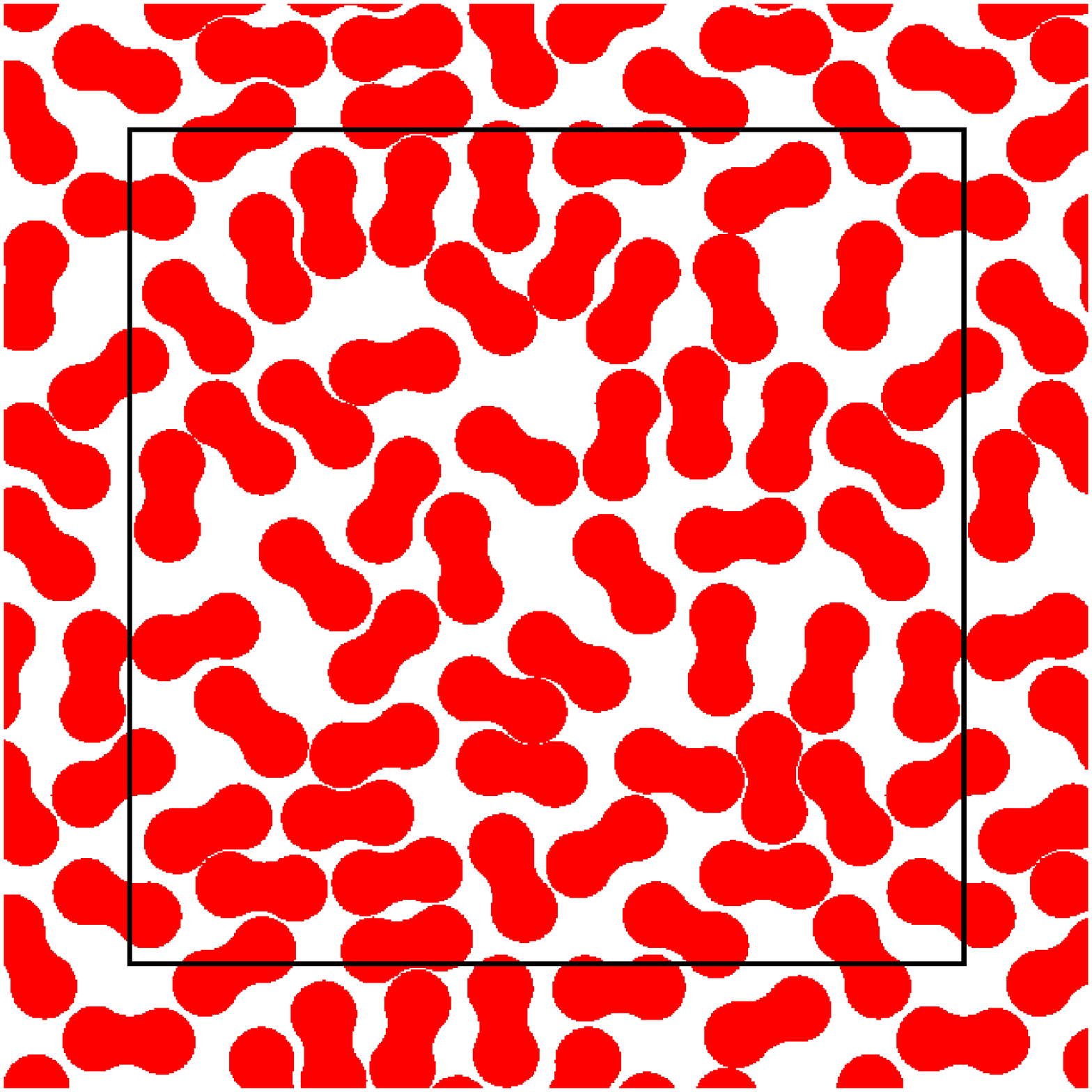}
\hspace{0.05\columnwidth}
\includegraphics[width=0.35\columnwidth]{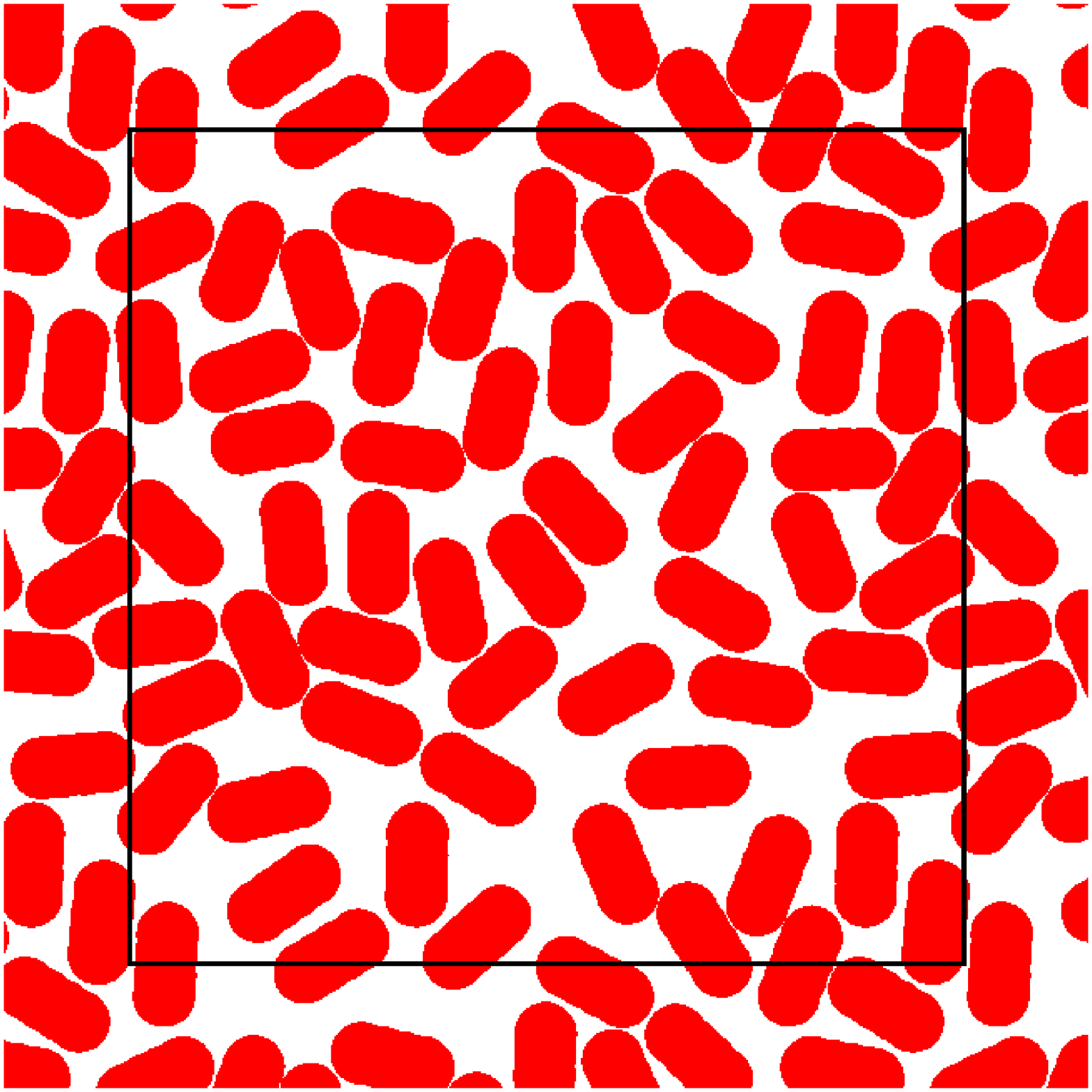}
}
\vspace{0.05\columnwidth}
\centerline{%
\includegraphics[width=0.35\columnwidth]{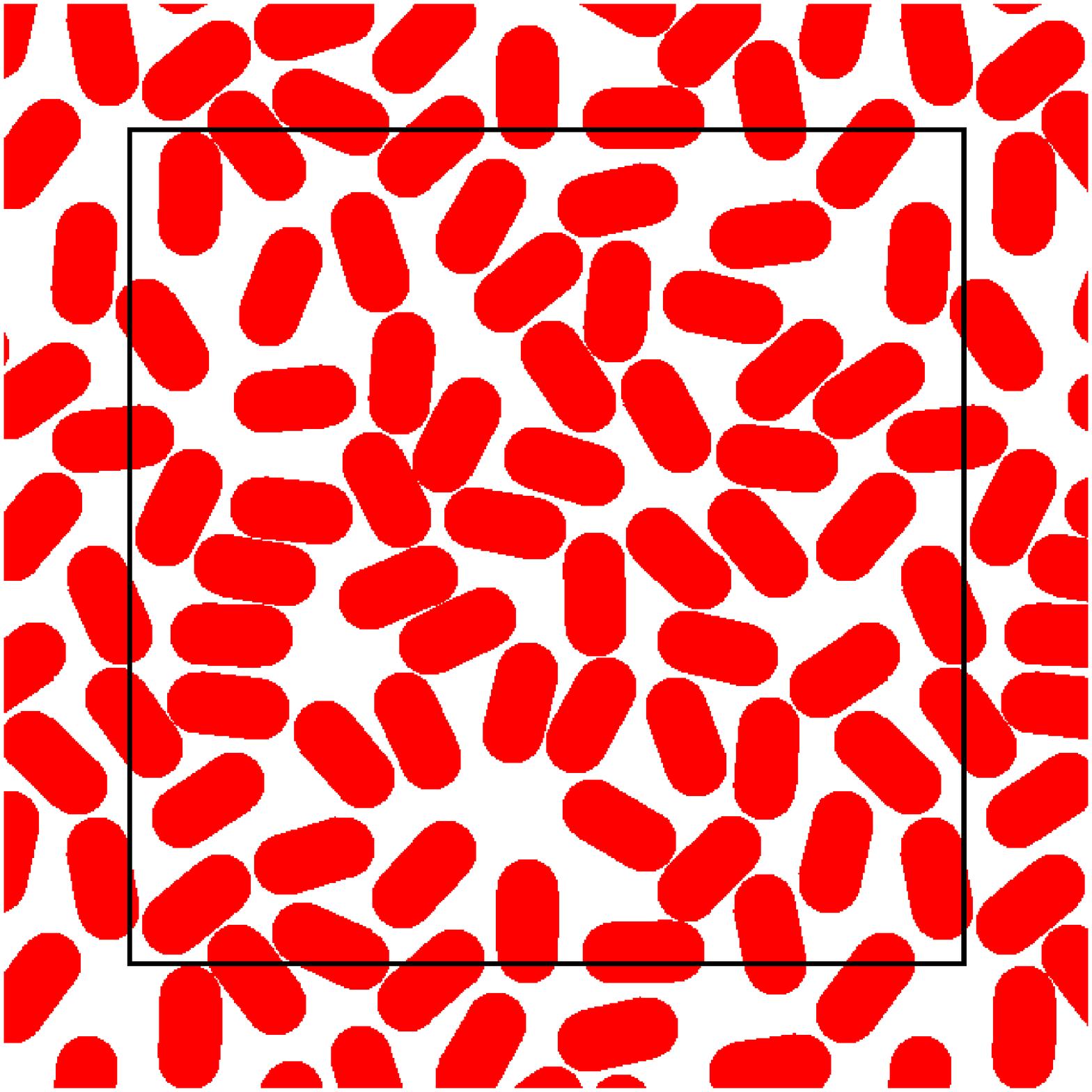}
\hspace{0.05\columnwidth}
\includegraphics[width=0.35\columnwidth]{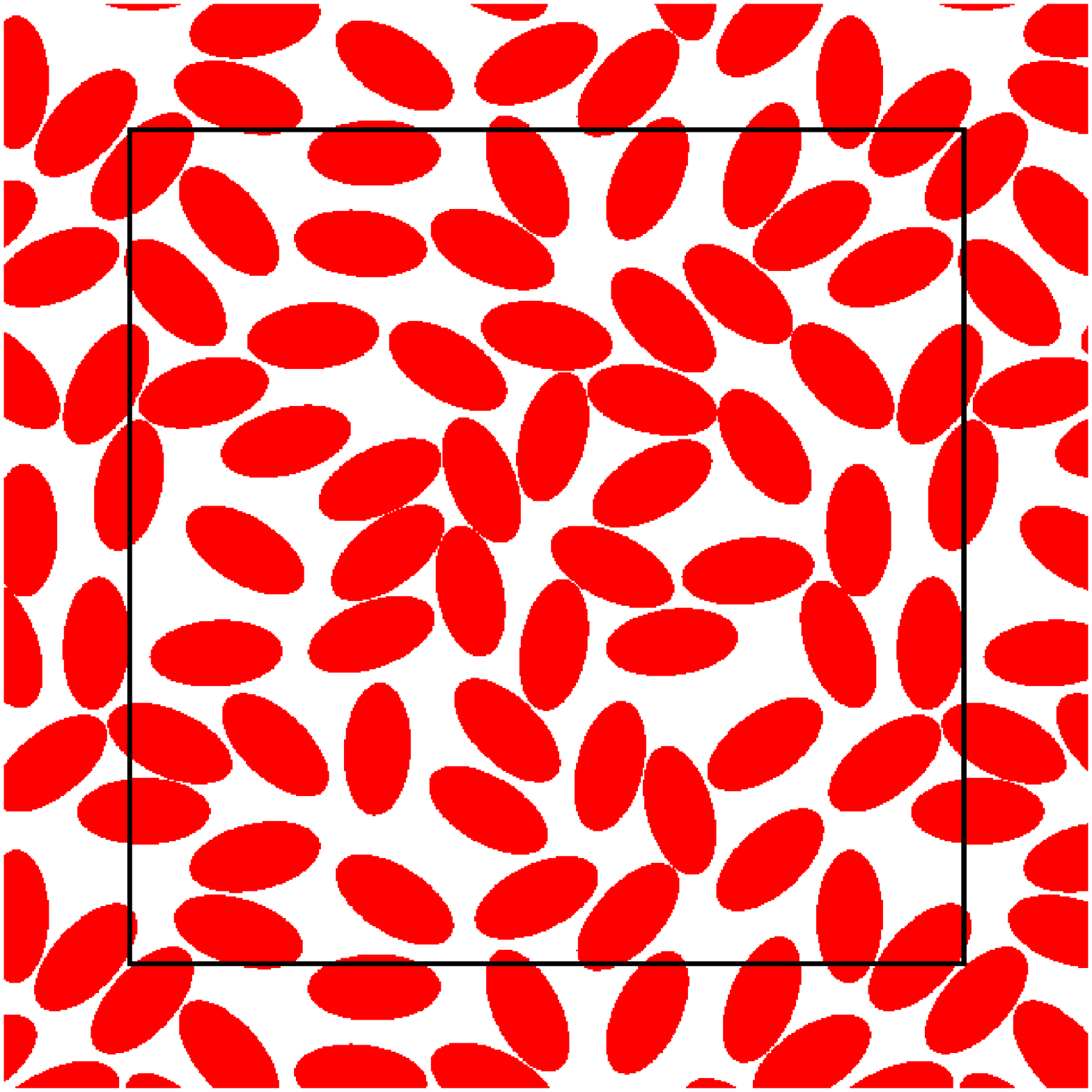}
}\caption{(Color online) Examples of obtained random packings of dimers, tetramers, spherocylinders and ellipses after $t=10^6$ iterations of RSA algorithm. The presented packings' sizes are $10 \times 10$ with periodic boundary conditions. The boundaries of the systems are indicated by black lines. The parameter $x$ equals $2.0$ for all four shapes.}
\label{fig:examples}
\end{figure}

Fragments of sample packings are presented in Fig.~\ref{fig:examples}.
Because the simulation is stopped after $t=10^6$, the resulting packings are most likely not saturated. To estimate the number of particles in saturated packing, the kinetics of the RSA are clarified in the following subsection. 
\subsection{RSA kinetics}
Asymptotically, for large enough time $t$, the kinetics of RSA is governed by the power law \cite{Pomeau1980,Swendsen1981}
\begin{equation}
\label{eq:fl}
 \theta(t) = \theta - A t^{-1/d}
\end{equation}
where $\theta \equiv \theta(t\to\infty)$, $A$ is a positive constant and $d$ depends on particle shape and properties of a surface on which particles are packed. For flat and homogeneous surfaces parameter $d$ can be interpreted as a number of degrees of freedom of a particle \cite{Ciesla2013pol, Hinrichsen1986}. Thus, for RSA of disks $d=2$, but for anisotropic particles $d=3$ because orientation of a particle gives an additional degree of freedom, even when the anisotropy is quite small \cite{Viot1992, Ciesla2014ring, Ciesla2014dim}. 
 
\begin{figure}[htb]
\centerline{%
\includegraphics[width=0.45\columnwidth]{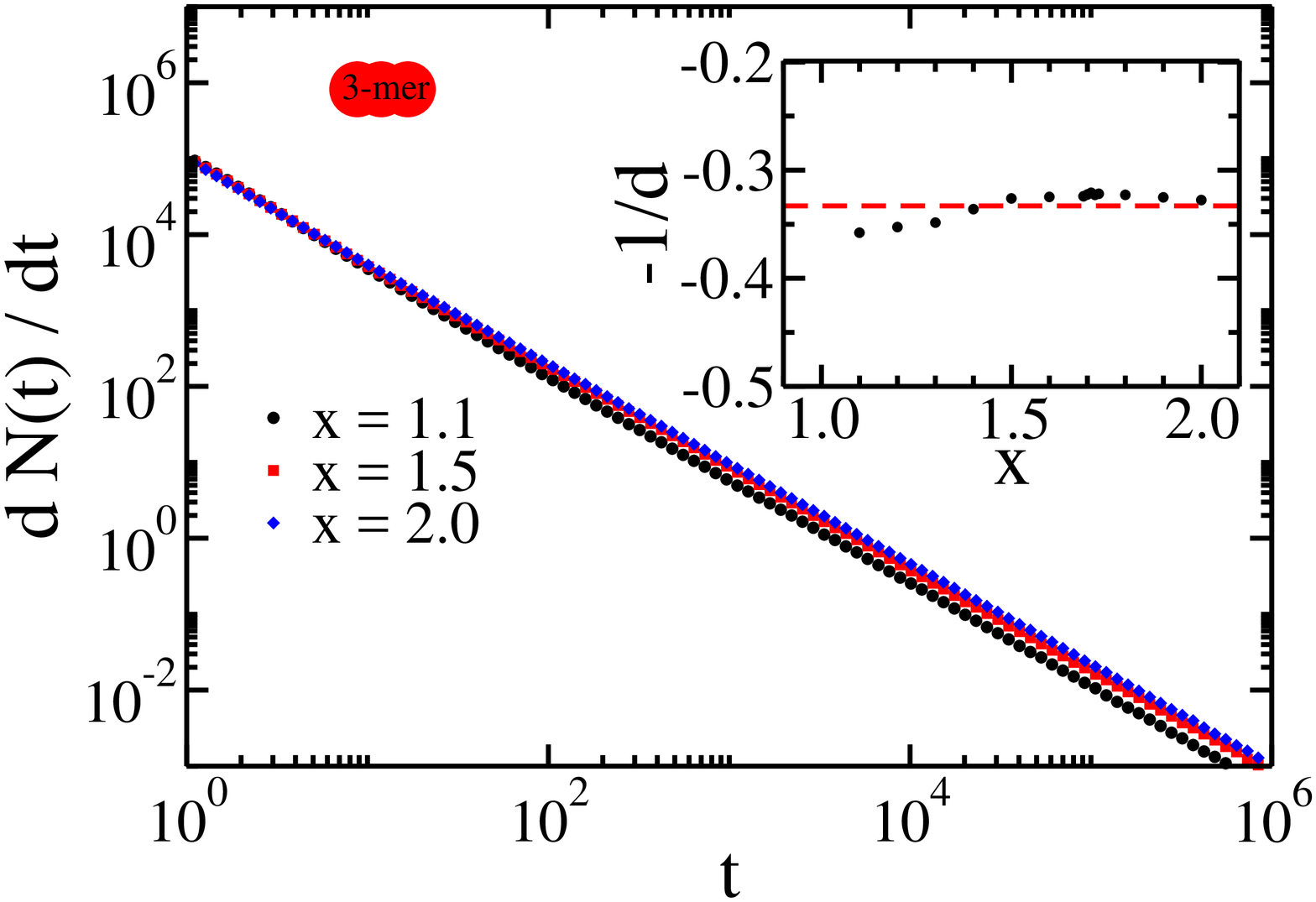}
\hspace{0.05\columnwidth}
\includegraphics[width=0.45\columnwidth]{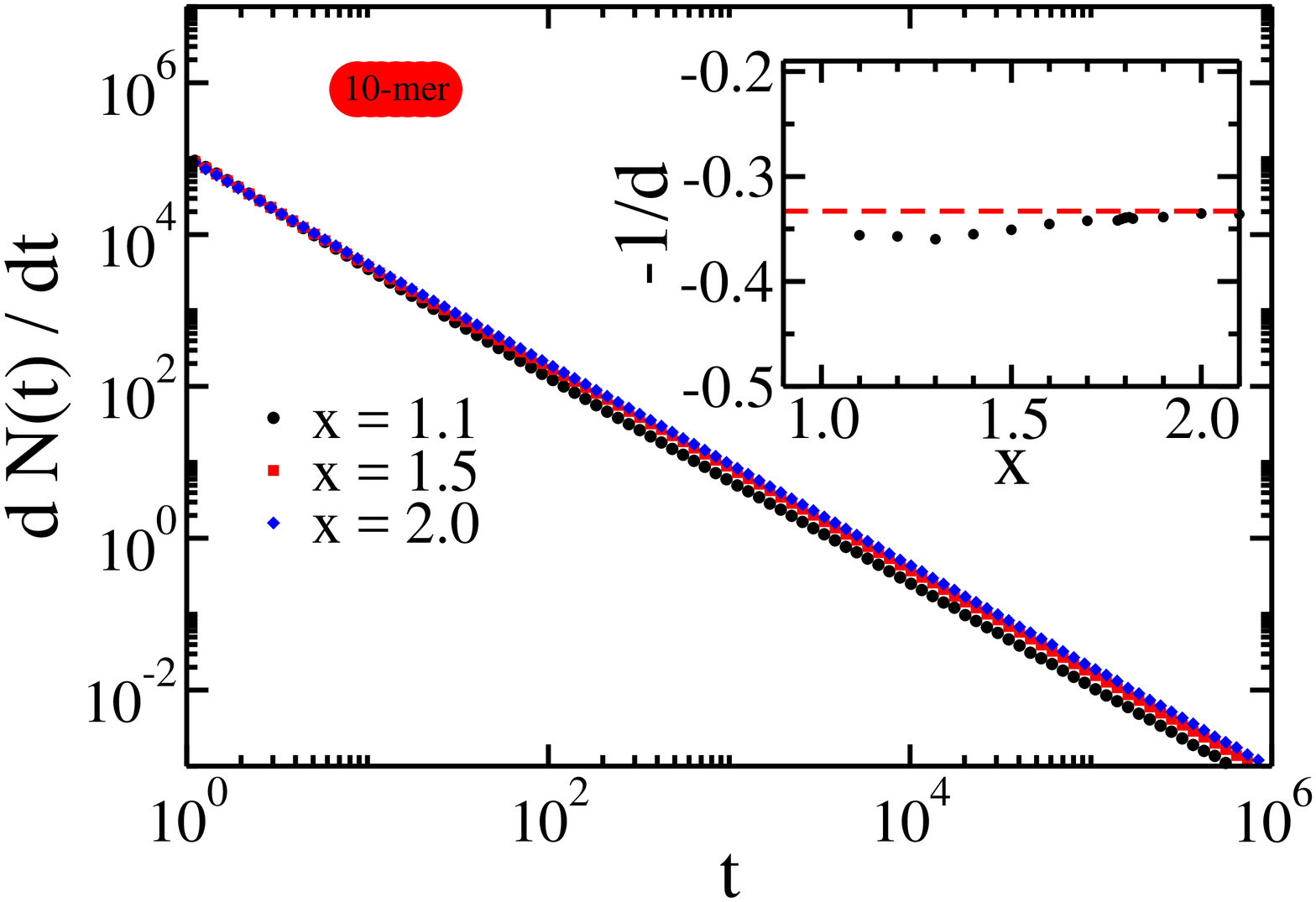}
}
\vspace{0.05\columnwidth}
\centerline{%
\includegraphics[width=0.45\columnwidth]{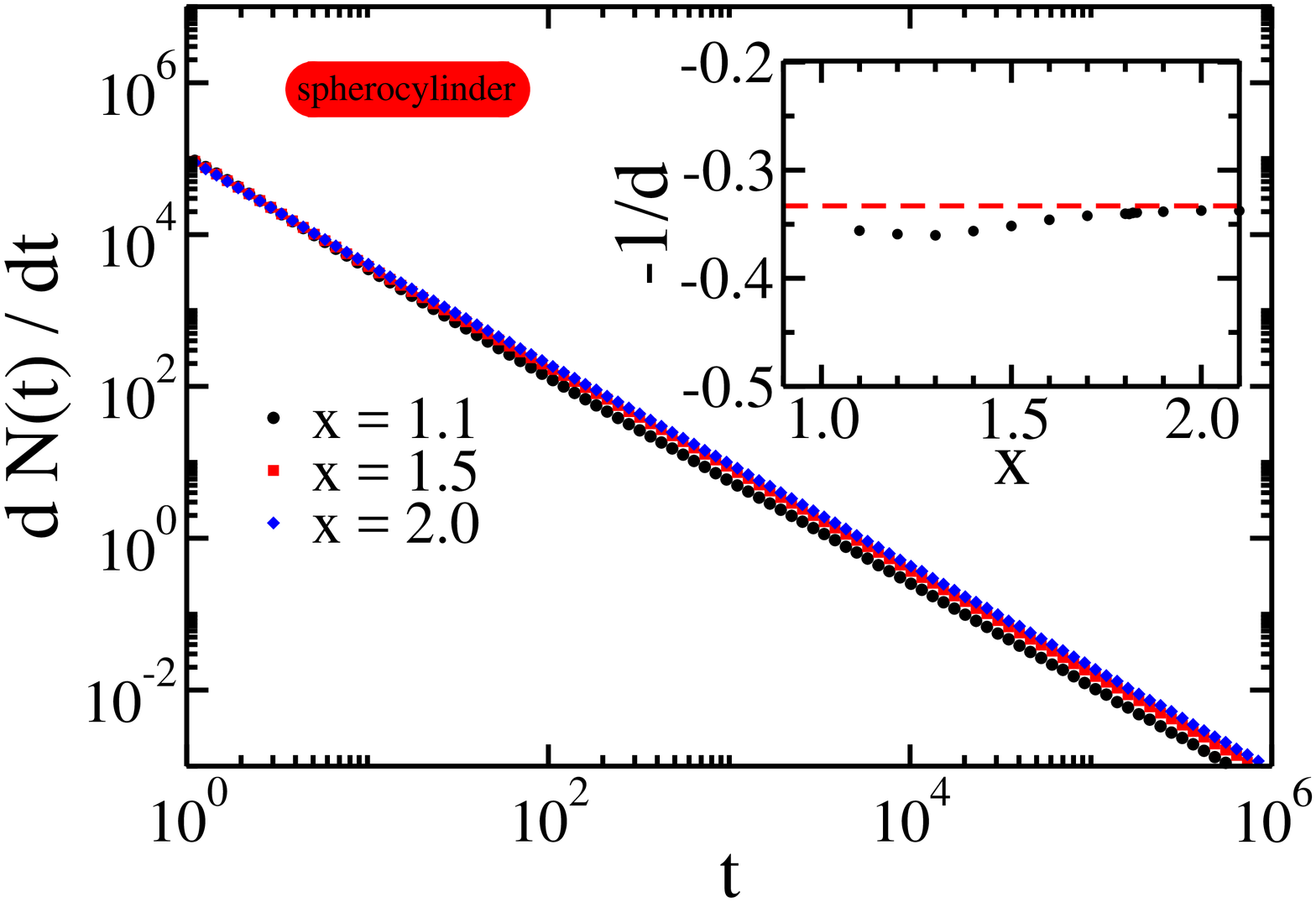}
\hspace{0.05\columnwidth}
\includegraphics[width=0.45\columnwidth]{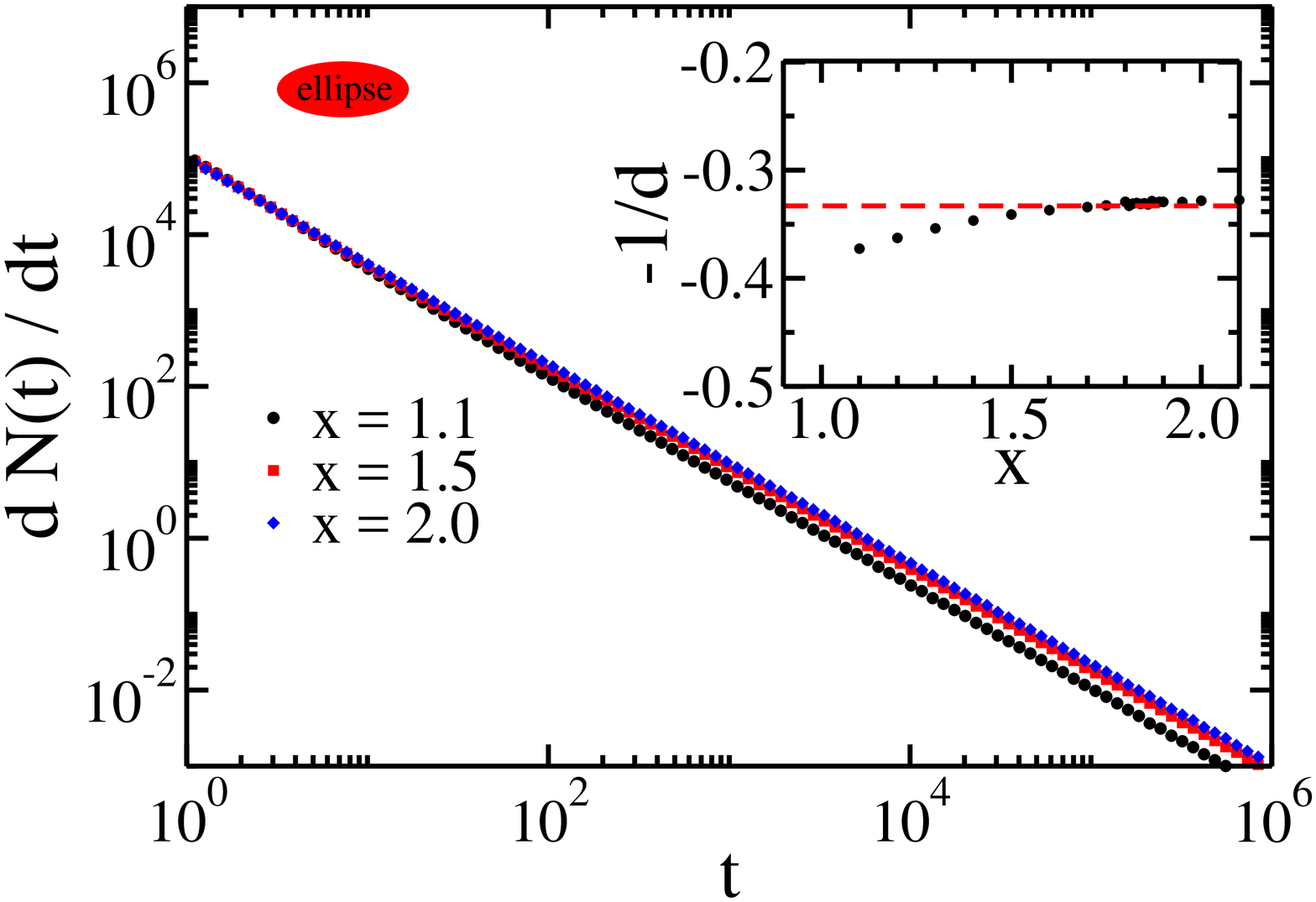}
}
\caption{(Color online) RSA kinetics for trimers, decamers, spherocylinders and ellipses. Main panels show number of added particles in a time unit (or the time derivative of $N$) versus time. Insets show the dependence on $x$ of the fitted exponent in Eq.~\ref{eq:fl}. Error bars are smaller than symbol sizes. Dashed lines correspond to $d=3$ degrees of freedom, which is characteristic of anisotropic molecules.}
\label{fig:kinetics}
\end{figure}

The examples of RSA kinetics of studied shapes are presented in Fig.~\ref{fig:kinetics}, where we plot $d N(t)/dt$ vs.\ $t$ on a log-log scale. 
The data for dimers has been presented in Ref. \ \cite{Ciesla2014dim}. Firstly, for all studied shapes, the numerical data in main panels in Fig.\ \ref{fig:kinetics} lie along straight lines, confirming that Eq.~(\ref{eq:fl}) is fulfilled. As expected the parameter $d$, obtained from fitting numerical data to the relation (\ref{eq:fl}) shown on insets in Fig.\ \ref{fig:kinetics}, is around $3$; it becomes slightly lower for small $x$, which agrees with previous observations  \cite{Viot1990, Ciesla2014dim}.
\subsection{Saturated random packing fractions}
\label{sec:packing}
The estimation of $\theta$ for finite-time simulations can be performed as follows. Having parameter $d$ and using a new variable $y = t^{-1/d}$, Eq.~(\ref{eq:fl}) can be converted to: $\theta(y) = \theta + A' y$. Thus, points $(\theta(y), y)$ measured during a simulation should lie along a straight line which crosses the axis $y=0$ at $\theta$.
The error of such $\theta$ estimation originates in error of the exponent $-1/d$, which in our simulation is at the order of $0.001$. The corresponding error of $\theta$ is in our case smaller than statistical error.

Another problem originates from the finite size of a system. According to Ref. \ \cite{Ciesla2016} for our setup this error should be comparable with the statistical error. Moreover, as we are mainly interested in comparing packing fraction of different shapes the results of such comparisons should not depend on system size, assuming that the system is big enough. 
The source of errors and its influence on obtained result are discussed in detail in Sec. \ref{sec:errors}.

\begin{figure}[htb]
\vspace{0.2in}
\centerline{%
\includegraphics[width=0.7\columnwidth]{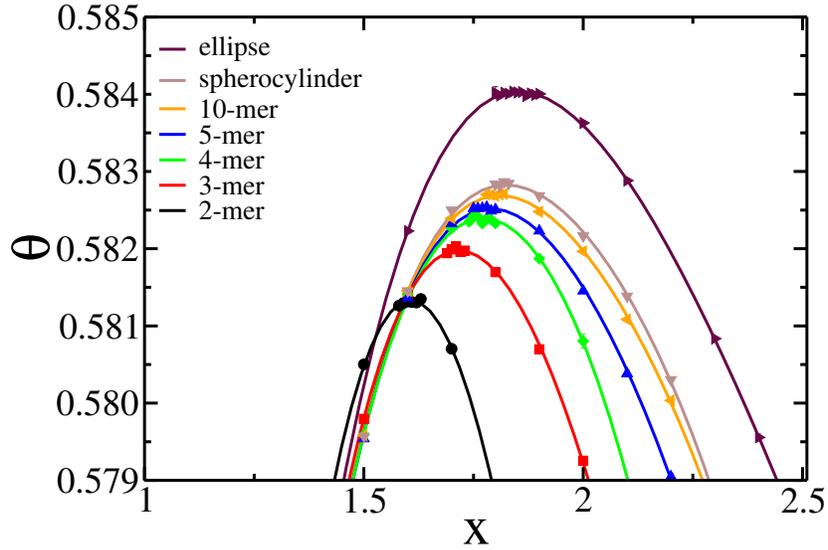}
}
\caption{(Color online) Saturated packing fraction dependence on parameter $x$ for all studied shapes. Dots represents data from numerical simulations. Bars corresponding to  statistical errors are smaller than symbol sizes. Solid lines are 4-th order polynomial fits: $\theta(x) = 0.29531 + 0.47397 \, x - 0.28573 \, x^2 + 0.075803 \, x^3 - 0.0076961 \, x^4$ for ellipses, $\theta(x) = 0.17371 + 0.75347 \, x - 0.5237 \, x^2 + 0.16448 \, x^3 - 0.019968 \, x^4$ for spherocylinders, $\theta(x) = 0.17846 + 0.73906 \, x - 0.50782 \, x^2 + 0.15703 \, x^3 - 0.018723 \, x^4$ for decamers, $\theta(x) = 0.18918 + 0.71263 \, x - 0.48458 \, x^2 + 0.14852 \, x^3 - 0.017676 \, x^4$ for pentamers, $\theta(x) = 0.14672 + 0.82962 \, x - 0.60441 \, x^2 + 0.20274 \, x^3 - 0.026843 \, x^4$ for tetramers, $\theta(x) = 0.2038 + 0.66753 \, x - 0.4367 \, x^2 + 0.12827 \, x^3 - 0.014935 \, x^4$ for trimers, and $\theta(x) = 0.19671 + 0.66073 \, x - 0.41212 \, x^2 + 0.11609 \, x^3 - 0.014196 \, x^4$ for dimers.}
\label{fig:q}
\end{figure}

The obtained packing fractions are presented in Fig.~\ref{fig:q}.%
For all shapes the maximal packing fraction is reached for $x \in [1.5; 2]$. It is similar to previous results for other shapes \cite{Ciesla2015} and confirms reasoning presented in \cite{Vigil1989} that for large $t$ anisotropy causes particles to align in parallel, which increases a packing fraction, but, on the other hand, at the beginning of RSA the anisotropic particle blocks significantly more space than a disk of the same area, which lowers the packing fraction. Thus, the optimum is reached for a small anisotropy. The data are fit to 4-th order polynomials, which allow us to accurately estimate an optimal anisotropy, and the value of the highest possible packing fraction. For convenience these data are collected together in the Table~\ref{tab:results}. Besides long-to-short axis ratio $x$ we used another measure of anisotropy, namely the shape factor, defined as \cite{Richard2001, Moucka2005}
\begin{equation}
\zeta = \frac{C^2}{4\pi},
\end{equation}
where $C$ is circumference of an object of unit surface area. Interestingly, the maximal packing fraction is reached for $\zeta = 1.136 \pm 0.011$ for all studied shapes, while the long-to-short axis ratio $x$ varies over a much wider relative range.

The statistical errors are of the order of $2.4 \cdot 10^{-5}$. Fluctuations of numerical values near the maxima (see Fig.\ref{fig:q}) suggest that the accuracy of the maximum coverage is a bit lower.
\begin{table}[htb]
\caption{\ Maximal possible saturated packing fractions and corresponding values of parameter $x$ for which they are reached. The error of $\theta$ does not exceed $10^{-4}$ (see Sec. \ref{sec:errors}). The error of $x$ corresponds to the width of the maximum of fitted function (see Fig.\ref{fig:q}), and is equal to $0.07$.}
\label{tab:results}
\begin{tabular*}{0.5\textwidth}{@{\extracolsep{\fill}}cccc}
\hline
shape & $x$ & $\zeta$ & $\theta$ \\
\hline
dimer           &  1.61 & 1.127 &  0.58132   \\
trimer          &  1.72 & 1.125 &  0.58200   \\
tetramer        &  1.77 & 1.130 &  0.58237   \\
pentamer        &  1.79 & 1.131 &  0.58249   \\
decamer         &  1.81 & 1.132 &  0.58269   \\
spherocylinder  &  1.82 & 1.133 &  0.58281   \\
ellipse         &  1.85 & 1.147 &  0.58405   \\
\hline
\end{tabular*}
\end{table}
It is worth commenting on the difference between the packing fraction of smoothed dimers obtained here and in Ref.\ \cite{Ciesla2015}, that is slightly larger than the error margin. The most probable cause of this discrepancy is the different boundary conditions used in the previous study (open boundaries), which most likely introduced a systematic error. A reliable comparison of packing fractions given by different shapes requires using the same boundary conditions for all of them.
\subsection{Structure of densest packing}
\label{sec:correlations}
\begin{figure}[htb]
\centerline{%
\includegraphics[width=0.7\columnwidth]{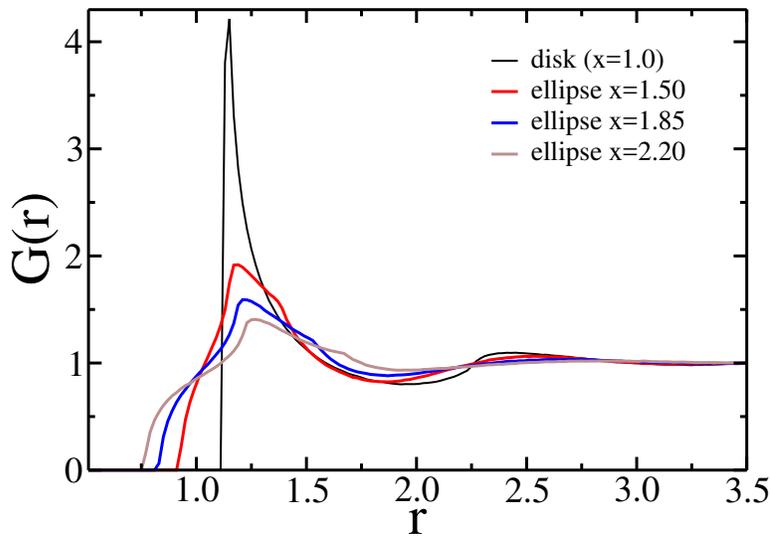}
}
\caption{(Color online) The density correlation function for ellipses of three different anisotropies. The black line is for saturated random packing of disks (ellipse with $x=1$) and takes a role of a reference frame. The distance is measured between particle centers. All shapes have unit surface area.}
\label{fig:correlations}
\end{figure}

Packing fraction contains information about mean density of shapes only. More details of packing structures can be obtained by studying correlation functions. Here we limit ourselves to the densest packing configuration of ellipses and two types of correlations. The first is the density correlation function which is proportional to the probability density function $p(r)$ of finding two particles, whose centers are separated by a distance $r$:
\begin{equation}
G(r) = \frac{p(r)}{\theta 2 \pi r}.
\end{equation}
The denominator is a normalization factor insuring that $G(r\to \infty) \to 1$.
The density correlation function for ellipses of different anisotropy is shown in Fig.\ \ref{fig:correlations}.
The presented correlation functions show behavior typical for packings build of anisotropic objects. They become non-zero for the closest possible distance between particle centers. Note that the larger anisotropy under a constant surface area condition implies a shorter possible distance between centers of neighboring objects, as they become thinner. That is why for larger $x$ the $G(r)$ start rising for smaller $r$. The growth is not as fast as for spheres because, due to different relative orientations, closest objects are at different distance from each other. This effect is stronger for larger anisotrpies. The maximum (near $r=1.2$) shifts to the larger distances with growth of anisotropy. The minimum is observed for $r \approx 2$. In general, the larger anisotropy, the smoother the density correlation function.

The second studied property of a packing structure is the local orientational ordering. Here we used the following definition of this parameter \cite{Ciesla2013pol}:
\begin{equation}
q(|\vec{r}|) = \left< 2\left[ \left< \left[ \hat{u}(\vec{x}) \cdot \hat{u}(\vec{x}+\vec{r}) \right]^2 \right>_r -\frac{1}{2} \right] \right>_x, 
\end{equation}
where $\hat{u}(\vec{x})$ is a unit vector along the long axis of a particle placed at point $\vec{x}$. The $\langle \cdot \rangle_r$ is an average over particles at a distance $r$, while $\langle \cdot \rangle_x$ is an average over different particle positions. The parameter $q$ is equal $1$ when particles are in parallel, and is equal $0$ when their orientations are random. The minimum value of $q=-1$ is possible if objects at a distance $r$ are perpendicular.

\begin{figure}[htb]
\centerline{%
\includegraphics[width=0.7\columnwidth]{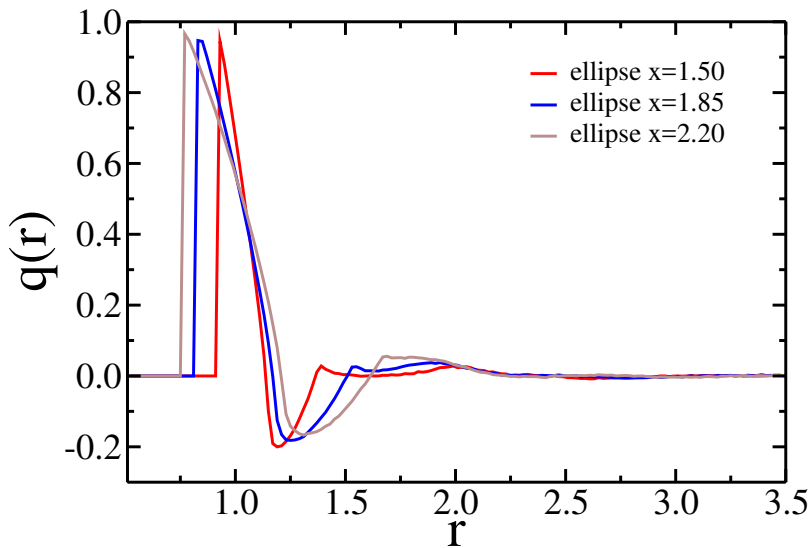}
}
\caption{(Color online) The propagation of the local orientational order inside a jammed sample of ellipses of three different anisotropies. The distance is measured between particle centers. All shapes have unit surface area.}
\label{fig:order}
\end{figure}

The local orientational ordering in packing is shown in Fig.\ \ref{fig:order}.
As expected, the closest possible placing of particles requires parallel alignment. The more interesting is the drop of $q(r)$ below zero for $r \approx 1.3$. To explain this, note that the minimum is near $(w+h)/2$ which is possible for a T-like configuration of neighboring shapes. Such configurations also prevents other particles to align in parallel, therefore the mean order is rather perpendicular than random. Particles at a distance $r>2$ are oriented randomly. 
\section{Estimation of measurement error}
\label{sec:errors}
As noted before, the setup of RSA algorithm was chosen accordingly to Ref.\ \cite{Ciesla2016} to ensure a level of numerical error of the average saturated random packing below $0.001$. To be sure that obtained results are precise enough we study in detail the case of RSA of ellipses of anisotropy $1.85$. 
In general, besides the statistical error of average packing fraction $\theta$, which was used in previous section, there are two sources of systematic error. They originate in finite number of RSA iterations and finite system size. The purpose of this section is to find out how these two sources affect the total error.

\begin{figure}[htb]
\centerline{%
\includegraphics[width=0.7\columnwidth]{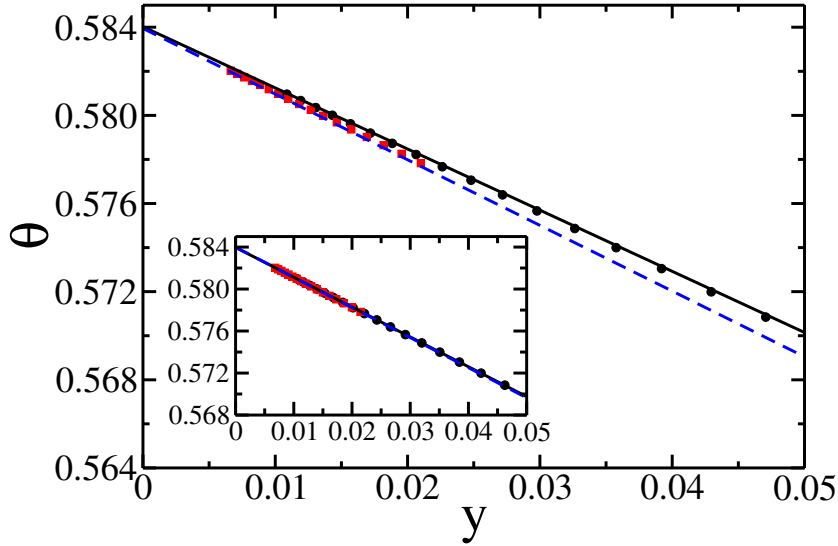}
}
\caption{(Color online) The dependence of the packing fraction of ellipses ($x=1.85$)  on $y=t^{-1/d}$. The fitted value of the parameter $d$ is $0.3314$ and $0.3352$, for simulations stopped at $t=10^6$ and $t=10^7$, respectively. Packing size $S=10^6$ was the same in both cases. Black dots and red squares are data obtained from simulations. Black solid and blue dashed lines are linear fits to these data. The obtained values of saturated packing fractions are $\theta = 0.58402 \pm 0.00002$ and $0.58395 \pm 0.00002$ for $t=10^6$ and $t=10^7$, respectively. Inset show the same data but instead of the fitted value of $d$, the theoretical one $d=1/3$ is used in both cases. Now the obtained values of saturated packing fractions are $\theta = 0.58396 \pm 0.00002$ and $0.58399 \pm 0.00002$ for $t=10^6$ and $t=10^7$, respectively.}
\label{fig:long}
\end{figure}

The statistical error depends on system size and number of independent packings. For $100$ independent square boxes of $S=10^6$ the standard deviation of average packing fraction $\theta$ is $2.4 \cdot 10^{-5}$. The influence of finite simulation time was estimated by generating independent packings up to dimensionless time $t=10^6$ and $10^7$. The results are presented in Fig.\ \ref{fig:long}.
The data were analyzed as described in Sec.\ \ref{sec:packing}. Slopes of both lines are slightly different because the measured value of $d$ is not the same in both the cases. However, both slopes agree within error limits. The difference between the obtained values of $\theta$ is $7 \cdot 10^{-5}$ and is approximately three times larger than the statistical error. 

\begin{figure}[htb]
\centerline{%
\includegraphics[width=0.7\columnwidth]{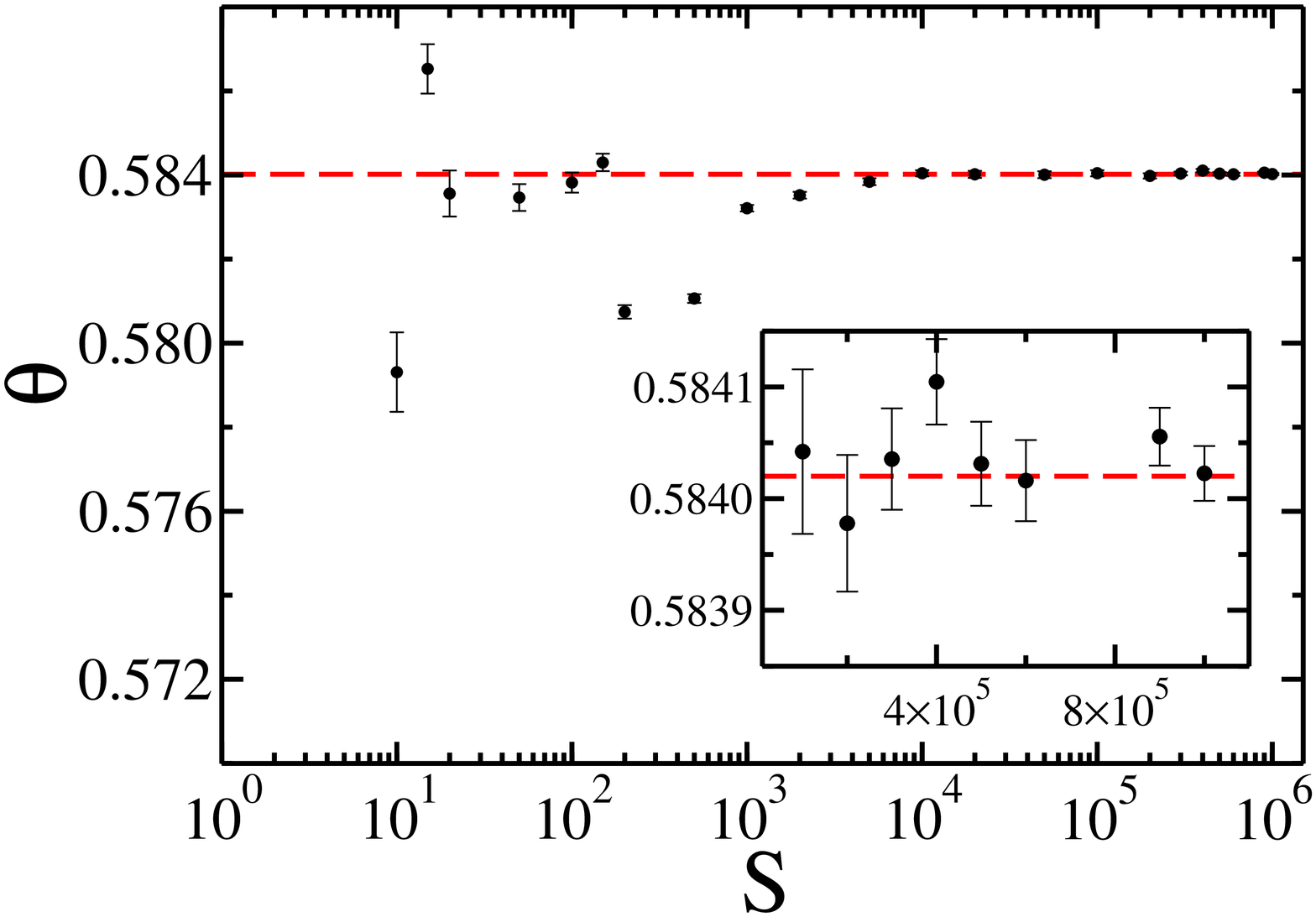}
}
\caption{(Color online) Dependence of the packing fraction of ellipses ($x=1.85$)  upon the size of surface $S = L^2$. Inset zooms the data for large $S$. Simulations were stopped at $t=10^6$. The red dashed line corresponds to $\theta = 0.58402$.}
\label{fig:size}
\end{figure}

The dependence of average packing fraction on system size is shown in Fig.\ \ref{fig:size}.
The results indicate that for $S=10^6$ finite-size effects are negligible in comparison with statistical errors (see Fig.\ \ref{fig:size} inset). It should be noted that in all studied cases the system size was quite large and periodic boundary conditions were implemented, thus significant finite size effects were not expected and it was confirmed by the data. Additional simulations performed for smaller packings suggest that finite-size effects can affect estimation of packing fraction for $S<10^4$. It is a little surprising as the density correlations in similar systems are superexponentialy damped and typically are not noticeable at distances $L \sim 10\ $\cite{Zhang2013, Bonnier1994}.
In summary, the main contribution to errors of presented results comes from finite simulation time and statistics. The total error of the average packing fraction of ellipses of $1.85$ anisotropy ratio is approximately $10^{-4}$ and is below needed level of accuracy. 

The error of an anisotropy $x$, for which the highest packing fraction occurs can be estimated as the half of the interval for which the fit $f(x)> \theta - \Delta\theta$ (see Fig.\ref{fig:q}). For studied shapes this condition gives $\Delta x \le 0.07$.
\section{Summary}
Several anisotropic and concave shapes of particles were analyzed in terms of maximal possible random packing fraction. It was found that the highest packing fraction is obtained for ellipses of long-to-short semi-axis ratio of $1.85 \pm 0.07$. The saturated random packing fraction for such a shape is $0.58405 \pm 0.0001$ which is higher than for smoothed $n$-mers ($0.58269 \pm 0.0001$ reached for anisotrpy $1.81 \pm 0.07$ for $n = 10$) and spherocylinders ($0.58281 \pm 0.0001$ reached for anisotropy $1.82 \pm 0.07$). Interestingly, the $n$-mers give smaller saturated packings than the spherocylinder, and one can see that with increasing $n$ the concave particles are closer to the behavior of the convex spherocylinder particle, being the extreme case. This outcome is somewhat expected, as the spherocylinder is the Minkowski sum (union of infinite number of disks along a finite line, see for instance \cite{Mulder2005, Rosso2006} and references therein) of the interval and a sphere (or disk in $2D$). Additionally, Table~\ref{tab:results} presents that the saturated packing fractions differ among those shapes less than the fraction of one percent, but at the same time corresponding anisotropies vary of the order of a dozen or so percent. In conclusion the presented results for the arrangements of shapes shown in Fig.~\ref{fig:shapes} indicate that particles giving similar maximal random coverage may significantly differ in their long-to-short axis ratio, although at the same time their shape factors $\zeta$ are nearly identical. 
\section*{Acknowledgments}
This research was carried out with the support of the Interdisciplinary Centre for Mathematical and Computational Modeling (ICM) at University of Warsaw under grant no.\ G-27-8. G.~Paj\c{a}k acknowledges support of Cracowian Consortium `$,\!,$Materia-Energia-Przysz\l{}o\'s\'c'' im.\ Mariana Smoluchowskiego' within the KNOW grant. 
\section*{Appendix}
\label{sec:appendix}
The area of linear polymer built of $k$ disks of unit radius, having overall long-to-short axis ratio $x$ is
\begin{equation}
  S_\mathrm{pol} = \pi +
  (k-1)\left( \frac{r_\mathrm{c}}{2} \sqrt{4-r_\mathrm{c}^2} + 2 \arcsin \frac{r_\mathrm{c}}{2} \right),
\end{equation}
where 
\begin{equation}
r_\mathrm{c} = \frac{x-1}{2(k-1)}
\end{equation}
is a distance between neighboring disks' centers.
The smoothed linear polymer area contains also $2(k-1)$ fragments of area:
\begin{equation}
S_\mathrm{ad} = \left\{
\begin{array}{c c}
 \frac{r_\mathrm{c}}{4} \left(\sqrt{16 - r_\mathrm{c}^2} -\sqrt{4 - r_\mathrm{c}^2} \right) - 
 \arcsin \frac{r_\mathrm{c}}{2} & 0 \leq r_\mathrm{c} < 2 \\
 \frac{r_\mathrm{c}}{4} \sqrt{16 - r_\mathrm{c}^2} -\frac{\pi}{2} & 2 \leq r_\mathrm{c} \le 2\sqrt{3}
\end{array}
\right.
\end{equation}
The circumference of such smoothed $k$-mer is: 
\begin{equation}
C_\mathrm{pol} = 2 [ 2\alpha + (2k-1)(\pi - 2 \alpha)],
\end{equation}
where $\alpha = \arccos r_c$ and $\alpha \in (0, \pi/2)$.

The area of a spherocylinder of height $h=2$ and long-to-short axis ratio $x$ is
\begin{equation}
S_\mathrm{sph} = \pi + 2(2x-2) .
\end{equation}
The circumference of such spherocylinder is: 
\begin{equation}
C_\mathrm{sph} = 2[\pi + (2x-2)].
\end{equation}

The area of an ellipse of short semi-axis equal to $1$ and long-to-short axis ratio $x$ is
\begin{equation}
S_\mathrm{ell} = x\pi .
\end{equation}
The circumference of an ellipse is known exactly in terms of the elliptic functions, but can be accurately estimated using the following relation due to Ramanujan \cite{Ramanujan1914}: 
\begin{equation}
C_\mathrm{ell} \approx \pi(x+1) \left[ \frac{3 (x-1)^2}{(x+1)^2\left(\sqrt{4-3\frac{(x-1)^2}{(x+1)^2}} + 10 \right)} +1 \right]
\end{equation}

To make the area equal to 1, we rescale the two dimensions by $1/\sqrt{S}$.
\bibliography{main.bib}
\end{document}